\documentclass[intlimits,twoside,a4paper]{article}

\usepackage[cp1251]{inputenc}

\usepackage[eqsecnum]{cmpj3}
\usepackage{array,multirow,makecell}
\usepackage{color}

\issue{2025}{28}{4}{43705}
\doinumber{10.5488/CMP.28.43705}

\title[Thermoelectric properties of PbSe$_{x}$Te$_{1-x}$]%
{A computational study of thermoelectric conversion in the PbSe$_{x}$Te$_{1-x}$ semiconductor alloys%
}
\author[M. Kaid Slimane, B. N. Brahmi, M. Bouchenaki, S. Bekhechi]
{M. Kaid Slimane\orcid{0009-0000-1011-0025}\thanks{Corresponding author: \email{kaidslimanem1999@gmail.com, meriem.kaidslimane@univ-tlemcen.dz}}, B. N. Brahmi\orcid{0000-0003-4576-0241},
M. Bouchenaki\orcid{0009-0009-9038-2092}, S. Bekhechi\orcid{0000-0002-2599-8554}  
\address{Theoretical Physics Laboratory, Faculty of Science, University Abou Bekr Belkaid, Tlemcen, Algeria }}


\date{Received April 21, 2025, in final form November 12, 2025}

\begin{document}
\maketitle

\begin{abstract}
The present theoretical study focuses on the structural, electronic and thermoelectric properties of PbTe, PbSe and their ternary alloys PbSe$_{x}$Te$_{1-x}$, using the density functional theory (DFT) by the full potential linearised augmented plane wave (FP-LAPW) method implemented in Wien2k code. Structural properties were performed by using the generalized gradient approximation of Perdew Burke and Ernzenhof (GGA-PBE) scheme. The results show that the calculated lattice parameters are in good agreement with theoretical data previously obtained. For electronic properties, we noticed that for all the compounds of PbSe$_{x}$Te$_{1-x}$, we have a direct band gap in L point. For thermoelectric properties, we used BoltzTraP2 code and Gibbs2 code. Our results show that the PbSe$_{x}$Te$_{1-x}$ compounds have reached a value of 2.55 for the figure of merit, which indicates that our material is a good thermoelectric candidate.

\printkeywords

\end{abstract}

\section{Introduction}
{Semiconductors are known in the field of scientific research for their electronic and optical properties~\cite{Kumar}, which leads to their widespread use across multiple sectors, such as in photovoltaic systems, electronic devices, and telecommunication equipment. Furthermore, in recent research, semiconductors have gained a new application focus. These materials have been studied in order to optimize their energy efficiency for thermoelectric devices \cite{And}.   }

{Thermoelectricity is a green technology, making it possible to generate an electric current from a flow of heat. The distinguishing feature of this modern technology is the energy converters, which are highly reliable, easy to use, emit no greenhouse gases and are clean systems, which underlines the importance of investigating this field and furthering scientific research \cite{Pan}. }

The capability of a material to be a good thermoelectric condidat is determined by the figure of merit \textit{ZT}, which is a factor linking four parameters: \textit{S}  (Seebeck coefficient), $\sigma$  (electrical conductivity), $\kappa$ (thermal conductivity) and \textit{T} (absolute temperature). The higher the \textit{ZT} value, the better the material is as a thermoelectric candidate. If a thermoelectric device has an average \textit{ZT} value greater than 1 in its operating temperature range, it is considered competitive and offers great potential for various technologies. The materials that make up those systems are chosen according to well-defined criteria: they must have a high Seebeck coefficient, high electrical conductivity and relatively low thermal conductivity. These characteristics are met by semiconductor materials \cite{Xwei,Jawad}, whose properties make them excellent candidates for thermoelectricity.

Recently, bismuth telluride $\textrm{Bi}_{2} \textrm{Te}_{3}$ \cite{Wit, Park}
has become the main material with identified thermoelectric properties. Today, a number of new avenues are being
explored, including skutterudites and lead chalcogenides. In this work, we referred to the data presented by Zair Asma et al. \cite{Zair}, who conducted a study on the thermoelectric properties of ternary alloys $\textrm{PbSe}_{1-x} \textrm{S}_{x}$ and obtained a \textit{ZT} value equal to~1. To investigate the different properties of ternary alloys PbSe$_{x}$Te$_{1-x}$, theoretical studies
 show that PbTe and PbSe, are stable in cubic structure \cite{Khan}
and cristallize in NaCl (rock salt) structure. Moreover, fundamental band gap for the PbTe and PbSe compounds is small and direct at the L-point \cite{Pan, Murt, Wei, Zhang, Bauer}, for thermoelectric properties. Khan et al. \cite{Khan}
have demonstrated that these compounds have nearly 0.76 value of \textit{ZT}, decreasing with temperature for PbTe due to large increscent in thermal conductivity with temperature as compared to its Seebeck coefficient and electric conductivity \cite{Khan}. Chen et al. \cite{Chen}
documented the achievement of employing atomic layer deposition (ALD) to synthesize nanolaminate structures consisting of alternating ternary PbSeTe layers and binary PbTe layers. This innovative approach aims at developing thermoelectric materials with high  \textit{ZT} values. To verify the efficiency of PbTe, PbSe, and their ternary alloys, we present a detailed study based on density functional theory to investigate the different structural properties of PbSe$_{x}$Te$_{1-x}$ alloys that crystallize in the NaCl structure. We also examine the electronic properties that determine the nature of the studied substances. Finally, the obtained results for the thermoelectric properties will be presented at the end of this work to assess the potential of these alloys and their efficiency for thermoelectric conversion.

\section{Computational details}
The calculations were performed by using the first principle calculation which is based on density functional theory (DFT) \cite{Shah, Louie, kovalen, bradji, kozina, sayah}, where Full Potential Linear Augmented Plane Wave FP-LAPW method~\cite{Petersen, adnane}
has been used, wich is implemented in the Wien2k code. The exchange-correlation potential was calculated by applying the generalized gradient approximation (GGA) using Perdew-Burke-Ernzerhof (PBE) \cite{Perdew}.

The lattice parameter, the bulk modulus and its first derivative are
extrapolated using the Murnaghan equation \cite{Murnaghan}. The electronic wave function, the charge density and the crystal potential were expanded in spherical harmonic indoors of the non-overlapping spheres surrounding the atomic sites (muffin-tin spheres) and by a plane-wave basis set in the interstitial regions, where the muffin-tin radius (RMT) was chosen as 2.5 for Pb, Te and Se.

A mesh of 47 and 216 special k-points for binary and ternary alloys respectively is used to analyze the influence of the number of $k$-points on the total energy in the irreducible wedge of the first Brillouin Zone (BZ) \cite{Boukhris}, the wave function cut-off was $R_{\rm MT}K_{\rm max}= 8$. The charge density was Fourier expanded in the interstitial region up to $G_{\rm max}$, Ry$^{1/2} =12$ {(Ry)}$^{1/2}$. In addition thermoelectric properties were performed by the use of the Boltzmann transport theory that is based on the second principle which is implemented in the BoltsTrap2 package \cite{Madsen}, and Gibbs2 code \cite{Otero,Otero2}
. Thus, we obtain the values of \textit{S}, $\sigma/\tau$, $\kappa_{e}/\tau$ and $\kappa_{l}$,{where $\tau$ is the relaxation time and the value used for all calculations is:
 $\tau$ = ${10}^{-14}$ s \cite{bouchenaki}}.

\section{Results and discussion}
\subsection{Structural properties}\label{sec3.1}
In this section, we study the effect of incorporating different concentrations of selenium (Se) into the binary alloy PbTe on the structural properties of the resulting alloys in NaCl structure. By using the method suggested by Agrawal et al. \cite{Agrawal}, we first calculated the lattice parameter $a_{0}$, 
bulk modulus $B$ and its first derivative $B'$ for the two binary compounds PbTe and PbSe. Then, we took a unit cell containing eight atomic sites, among which four sites are occupied by the element Pb and the remaining four sites by Te. After that we replaced each time an atom of Te by an atom of Se. {This result consolidates the experimental and theoretical studies reported on semiconductor alloys showing the violation of the Virtual Cristal Approximation (VCA).} 
This will be done in three steps in order to obtain the three ternary alloys  $\textrm{PbSe}_{0.25} \textrm{Te}_{0.75}$,  $\textrm{PbSe}_{0.5} \textrm{Te}_{0.5}$,  $\textrm{PbSe}_{0.75} \textrm{Te}_{0.25}$ with concentrations of Se: \textit{x} = 0, 0.25, 0.5, 0.75 and~1. We evaluated the lattice parameters $a_{0}$, bulk modulus $B$ and its first derivative $B'$ {at 0 K \cite{Blaha} } by fitting the unit cell energy versus its volume using the Murnaghan equation \cite{Murnaghan}. We have estimated the value of the bulk modulus from the $E(V)$ curve, where $B$ represents the minimum of it. The structural parameters are calculated at the equilibrium state which is given by the point with the lowest energy. In table \ref{table1}, we summarize the experimental data of our alloys, {that have been mesured at 300~K \cite{Zhang}} 
for the three parameters, as well as the results obtained using the GGA-PBE method, and those from other calculations for comparison.

\begin{table}[!t]
    \centering
    \caption{Lattice parameter $a_{0}$, bulk modulus $B$ and its first derivative $B'$.}
    \vspace{0.2cm}
    \begin{tabular}{ccccc}
    \hline
    Alloy & Method & $a_0$ (\AA)   & $B$ (GPa) & $B'$  \\ 
    \hline
     \multirow{4}{5em}{PbTe}  & Exp & $6.462^{a}$, $6.454^{b}$ & $39.8^{a}$ & - \\
      & Our work  & $6.5667$ & $39.3204$ & $4.8189$  \\
      & Other work &  $6.38^{a}$, $6.565^{b}$,  & $38.04^{a}$, $44.4^{c}$,  & $5^{c}$ \\
      & & $6.42^{c}$, $6.56^{d}$ & $38.406^{d}$ & \\
      \hline
      \multirow{3}{6em}{$\textrm{PbSe}_{0.25} \textrm{Te}_{0.75}$}  & VCA & $6.3775$ & - & - \\
      & Our work  & $6.4893$ & $39.4483$ & $4.6473$  \\
      & Other work &  $6.36^{c}$, $6.487^{d}$  & $46.6^{c}$, $37.791^{d}$  & $5^{c}$ \\
      \hline
      \multirow{3}{6em}{$\textrm{PbSe}_{0.5} \textrm{Te}_{0.5}$}  & VCA & $6.293$ & - & - \\
      & Our work  & $6.4072$ & $40.5282$ & $4.3565$  \\
      & Other work &  $6.27^{c}$, $6.406^{d}$  & $49.26^{c}$, $41.101^{d}$  & $5^{c}$ \\
      \hline
      \multirow{3}{6em}{$\textrm{PbSe}_{0.75} \textrm{Te}_{0.25}$}  & VCA & $6.2085$ & - & - \\
      & Our work  & $6.3164$ & $42.9262$ & $4.4804$  \\
      & Other work &  $6.199^{c}$, $6.32^{d}$  & $51.67^{c}$, $44.489^{d}$  & $5^{c}$ \\
      \hline
      \multirow{5}{6em}{PbSe}  & Exp & $6.124^{a,b}$ & $54.1^{a}$ & - \\
      & Our work  & $6.2111$ & $45.5375$ & $4.9284$  \\
      & Other work &  $6.036^{a}$, $6.196^{b}$,  & $49.1^{b}$, $55.75^{c}$,  & $5^{c}$ \\
      & & $6.095^{c}$, $6.21^{d}$, & $49.187^{d}$, & \\
       & & $6.2151^{e}$ & $48.661^{e}$ & \\
      \hline
      \multicolumn{5}{l}{a: GGA \cite{Zhang}
      , b: GGA \cite{Lach}
      , c: GGA \cite{Murt}
      , d: GGA \cite{Boukhris}
      , e: GGA \cite{Zair}
      }
    \end{tabular}
    \label{table1}
\end{table}

The results obtained in table \ref{table1} show a discrepancy between the calculated value of the lattice parameter using the GGA-PBE method and the experimental value available for the binary compounds PbTe and PbSe which is attributed to the fact that the method used is known to provide an overestimated result compared to the exact experimental value, as mentioned by Brahmi et al. \cite{Brahmi}.

Furthermore, regarding the calculated values for the bulk modulus shown in table \ref{table1}, we observe a difference between the calculated values and the experimental ones, while they are in agreement with the results obtained by other calculations.

\begin{figure}[!t]
\centering
\includegraphics[width=1\textwidth]{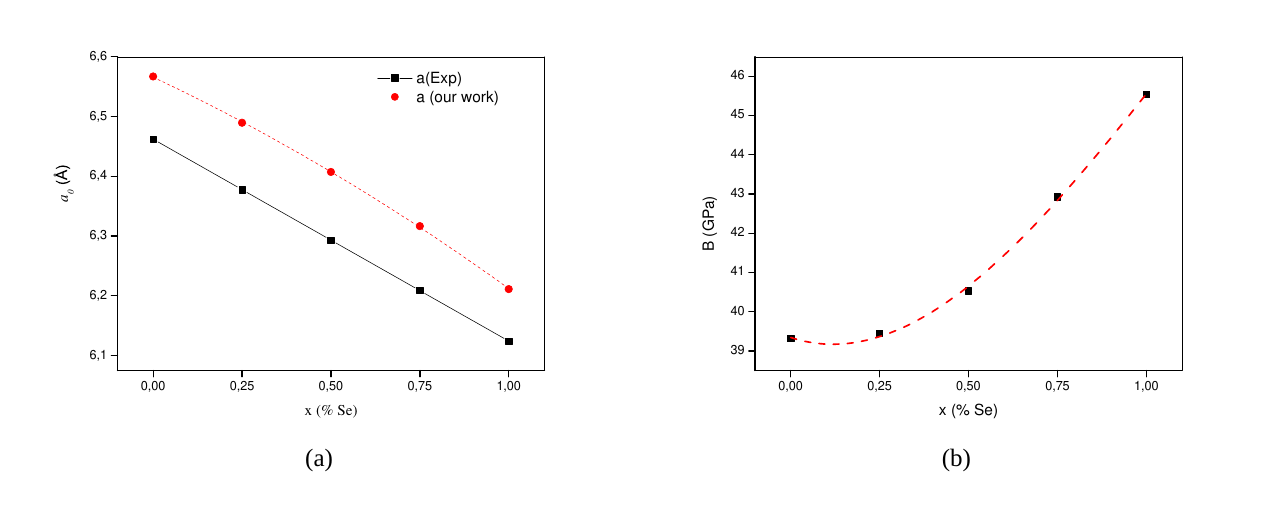}
\caption{(Colour online) Variation of (a) lattice parameter $a_0$ (b) bulk modulus $B$ versus concentration \textit{x} of Se.}
\label{fig1}
\end{figure}

In figure~\ref{fig1} (a), we plotted the variation of the lattice parameter $a_{0}$ as a function of the concentration \textit{x} of the Se element in the PbSe$_{x}$Te$_{1-x}$ alloys. Despite the difference between the experimental and calculated values, the resulting curve shows a linear variation with a smooth fit to the calculated values, which is in good agreement with Vegard's prediction \cite{Vegard}. The fitting of these points is represented in the figure and given is by the equation \eqref{eqt1} of order 2:

\begin{equation}\label{eqt1}
    y=6.5658-0.2801x-0.0735x^{2}.
\end{equation}

As we can notice, the fit is of order 2, and the linearity offset is very small (bowing parameter equal to $-0.0735$~\AA), which confirms that the results obtained are in good agreement with Vegard's prediction.

In  figure~\ref{fig1} 
(b), we plotted the variation of bulk modulus B versus concentration \textit{x} of Se. We note that this parameter increases from 39.32 GPa to 45.53 GPa, with a nonlinear variation, the fitting of these points is shown in the figure, and is given by equation \eqref{eqt2}:

\begin{equation}\label{eqt2}
    y= 39.3414-2.9355x+ 13.0923x^{2} -3.9397x^{3}.
\end{equation}

A second time we observe a slight non-linearity of the values for the bulk modulus, with a bowing parameter equal to $-3.9397$~GPa, which is small compared to the values we obtained, with a fit of order~3.

\subsection{Electronic properties}

The electronic properties of a material make it possible to identify its nature, and also to optimize its use in the appropriate field. For example, semiconductors are used in various applications such as photovoltaic cells, transistors, lasers and laser diodes \cite{Chanda}. In this section, we present the calculated values of the energy gaps using the GGA-PBE method as well as the variation of the energy gap versus concentration \textit{x} of Se.

In table \ref{table2}, we present the calculated values of the energy band gap for our PbSe$_{x}$Te$_{1-x}$ alloys using GGA-PBE scheme, along with experimental data { at very low temperatures \cite{Ravindra} } and results obtained from other studies.

\begin{table}[!t]
    \centering
    \caption{Calculated band gap for PbSe$_{x}$Te$_{1-x}$ alloys.}
    \vspace{0.2cm}
    \begin{tabular}{cccc}
    \hline
    \multirow{3}{*}{Alloy}   & \multicolumn{3}{c}{$E_g$ (eV)}  \\ 
    \cline{2-4}
     & This work  & Other work  & \multirow{2}{*}{Exp}  \\
     \cline{2-3}
     & GGA-PBE  & GGA  & \\
     \hline
     PbTe & 0.824  & 0.7$^a$ & 0.22$^b$ \\
     \hline
     $\textrm{PbSe}_{0.25} \textrm{Te}_{0.75}$ & 0.197 & 0.12$^a$ & -\\
     \hline
     $\textrm{PbSe}_{0.5} \textrm{Te}_{0.5}$ & 0.229  & 0.09$^a$ & -\\
     \hline
     $\textrm{PbSe}_{0.75} \textrm{Te}_{0.25}$ & 0.295   & 0.13$^a$ & -\\
     \hline
     PbSe & 0.420  & 0.24$^a$ &  0.17$^b$ \\
     \hline
     \multicolumn{4}{l}{a: \cite{Murt}
     , b: \cite{Ravindra}
     }
    \end{tabular}
    \label{table2}
\end{table}

We can notice that the calculated values for the binary compounds PbSe and PbTe are overestimated compared to experimental data. This significant difference between the calculated band gap energy using the DFT formalism and the experimental value is primarily attributed to an intrinsic characteristic of DFT. This occurs because DFT is a theory based on the fundamental state and is not
suitable for describing the properties of excited states as it has been mentioned by Boukhris et~al.~\cite{Boukhris}. However, our results are in agreement with those obtained from other studies.

\begin{figure}[!t]
\centering
\includegraphics[width=0.6\textwidth]{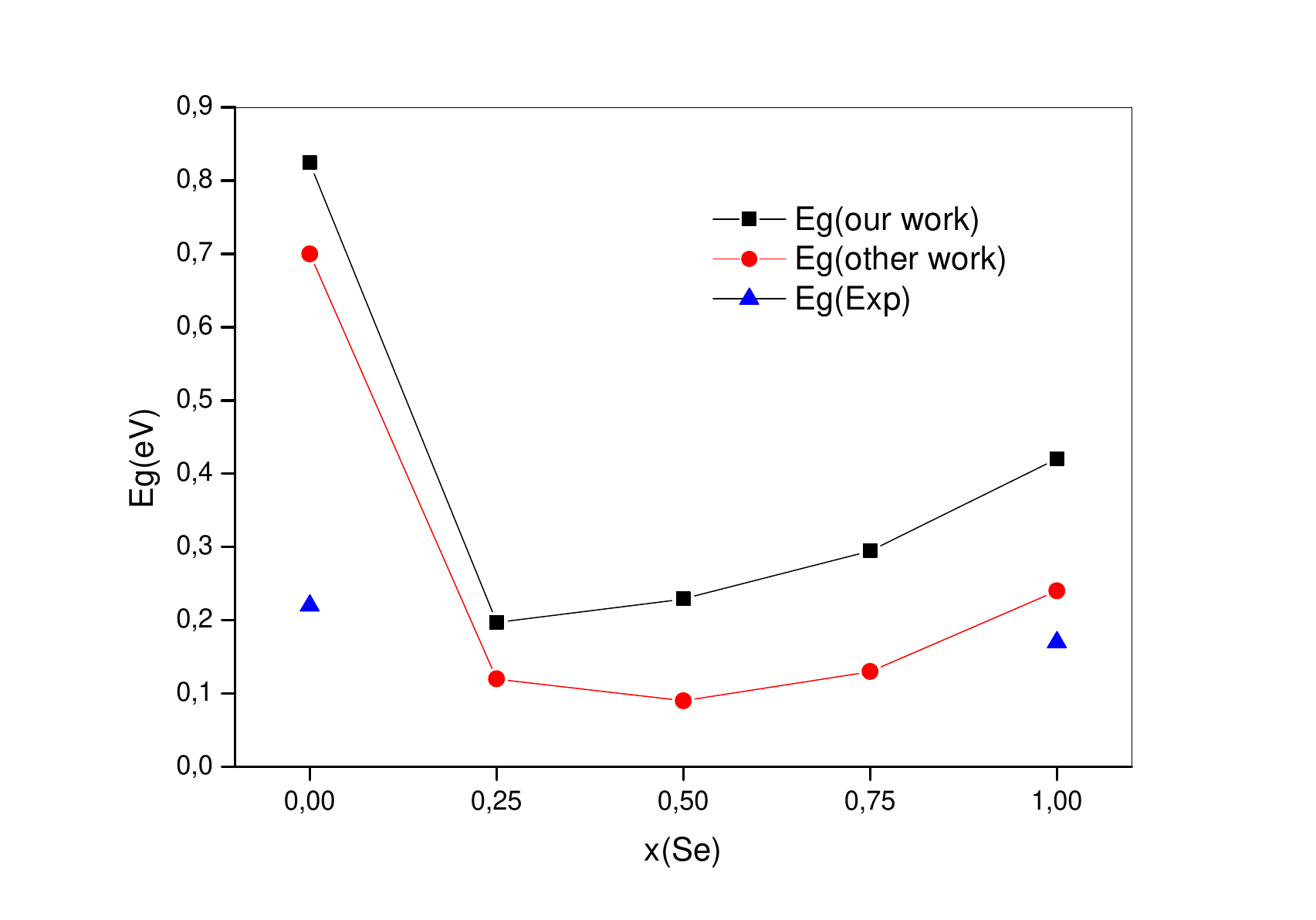}
\caption{(Colour online) Composition dependence of the calculated band gap for PbSe$_{x}$Te$_{1-x}$.}
\label{fig2}
\end{figure}

To better visualize the dependence of the gap value on concentration, we plotted the variation of the energy gap as a function of the composition \textit{x} in  figure~\ref{fig2}.
In figure \ref{fig2}, a nonlinear variation of the energy band gap versus the Se concentration \textit{x} in the PbSe$_{x}$Te$_{1-x}$ alloy is observed. The fitting of these points is given by equation \eqref{eqt3}:

\begin{equation}\label{eqt3}
    y=0.842-5.3093x + 15.0388x^2 -17.0666x^3 +6.9333x^4.
\end{equation}

This nonlinearity has also been observed by Murtaza et al. \cite{Murt}, Boukhris et al. \cite{Boukhris}, and Labidi et~al.~\cite{Labidi}. It is noticed that the incorporation of 25\% of Se leads to a decrease in the gap value. Afterward, the band gap value increases until reaching the value of 0.42 eV. Previous studies have demonstrated that the energy gap exhibits a strong dependence on the composition of the alloys, which is attributed to the lattice parameter and the electronegativity difference of the parent atoms PbTe and PbSe \cite{Boukhris, Charifi, Van}.

\subsection{Thermoelectric properties}\label{sec3.3}

A good thermoelectric material must have a high Seebeck coefficient and a high electrical conductivity, as well as a low thermal conductivity \cite{Charifi}. Consequently, the figure of merit \textit{ZT}, which represents the thermoelectric capacity of the material, links these three parameters through equation \eqref{eqt4}. The higher the value of \textit{ZT}, the better the material is as a thermoelectric candidate

\begin{equation}\label{eqt4}
    ZT=\frac{S^2\sigma}{\kappa}T .
\end{equation}

{To study the capability of materials to convert the heat flow into electrical current, we need to calculate the four fundamental parameters of thermoelectricity. To do this, we used the {BoltzTrap2} \cite{Madsen} code integrated into the {Wien2k} \cite{Blaha} interface. We varied the temperature in the range (120 K, 1200 K) and calculated the Seebeck coefficient \textit{S}, electrical conductivity $\sigma$, thermal conductivity $\kappa$, and figure of merit \textit{ZT} for PbSe$_{x}$Te$_{1-x}$ alloys.
For thermal conductivity, we used \textit{Gibbs2} \cite{Otero,Otero2} software to perform the necessary calculations \cite{bouchenaki}.
}

For this purpose, the chemical potential values that we took are the closest to Fermi level, and are given in table \ref{table3} :

\begin{table}[!t]
    \centering
    \caption{Chemical potential of the  alloys PbSe$_{x}$Te$_{1-x}$.}
    	\vspace{0.2cm}
    \begin{tabular}{cccccc}
    \hline
     Compound  & PbTe  & $\textrm{PbSe}_{0.25} \textrm{Te}_{0.75}$ & $\textrm{PbSe}_{0.5} \textrm{Te}_{0.5}$ & $\textrm{PbSe}_{0.75} \textrm{Te}_{0.25}$ & PbSe  \\
     \hline
     Chemical  & \multirow{2}{*}{0.324191}  &  \multirow{2}{*}{0.326667} &  \multirow{2}{*}{0.325675} &  \multirow{2}{*}{0.326081} &  \multirow{2}{*}{0.327519} \\
     potential $\mu_0$ (Ry) & & & & & \\
     \hline
    \end{tabular}
    \label{table3}
\end{table}

\subsubsection{Seebeck coefficient}

\begin{figure}[!t]
\centering
\includegraphics[width=0.7\textwidth]{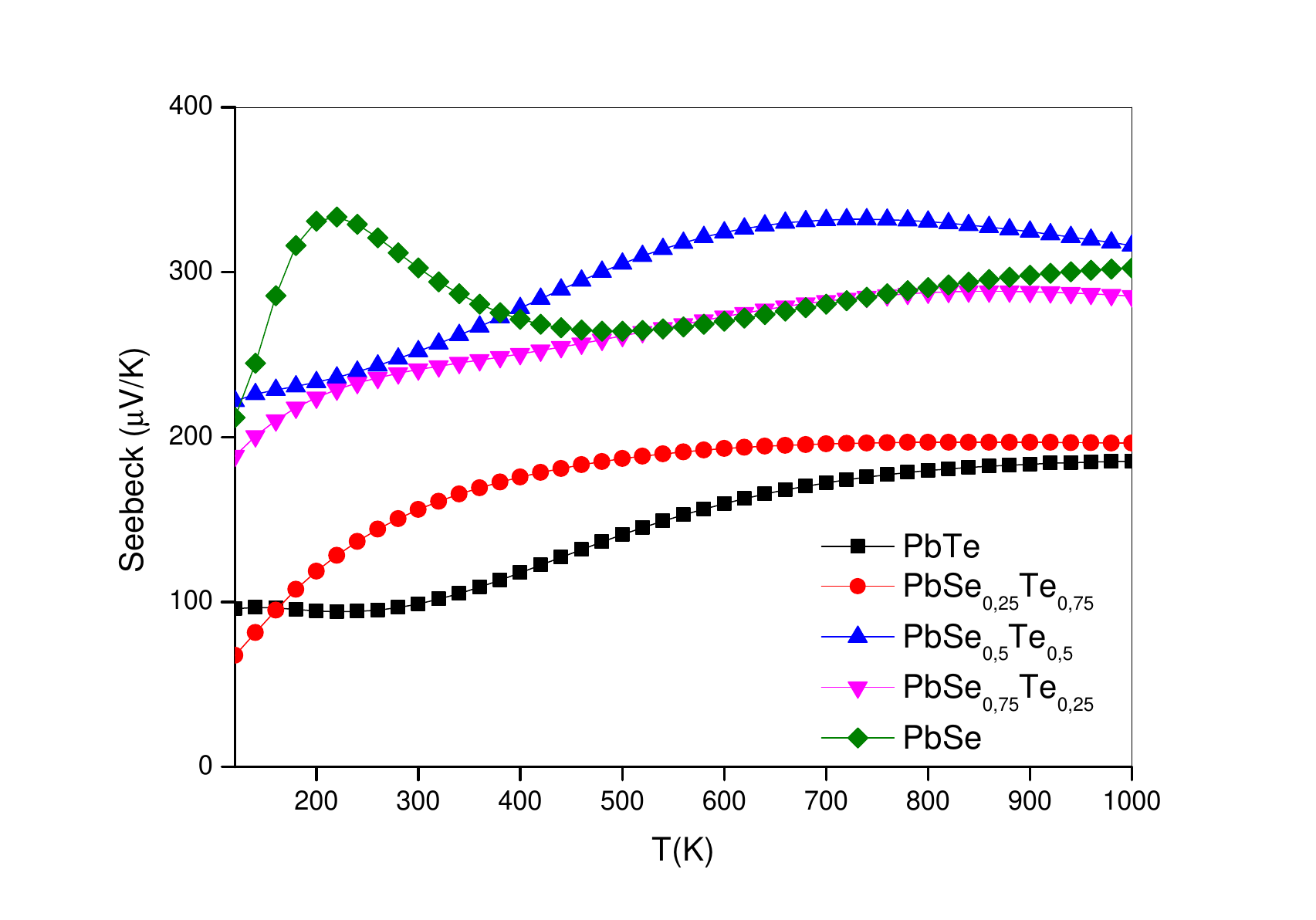}
\caption{(Colour online) Variation of Seebeck coefficient versus temperature for PbSe$_{x}$Te$_{1-x}$alloys.}
\label{fig3}
\end{figure}

The results obtained after calculating the Seebeck coefficient over a temperature range from 120~K to 1000~K for the five compounds are plotted in figure \ref{fig3}. It can be observed that the values of this coefficient are of the order of microvolts per Kelvin. Furthermore, it is noted that for the two compounds PbTe and
$\textrm{PbSe}_{0.25} \textrm{Te}_{0.75}$, the Seebeck coefficient increases and converges to a value of 185 \textmu{}V/K and 195 \textmu{}V/K respectively, in the range of high temperatures. For the $\textrm{PbSe}_{0.5} \textrm{Te}_{0.5}$, $\textrm{PbSe}_{0.75} \textrm{Te}_{0.25}$, and PbSe alloys, the Seebeck coefficient decreases and converges to a value of 316 \textmu{}V/K, 285 \textmu{}V/K and 300 \textmu{}V/K respectively, in the range of high temperatures.

It should be noted that semiconductors are of low charge carrier concentration and have a large effective mass, which explains the magnitude of the values obtained for the Seebeck coefficient \cite{Zair, Khan}.

{A positive Seebeck coefficient indicates that the majority charge carriers responsible for electrical and thermal transport are holes (positive charge, hence the name \textit{p}-type for positive), in this case, the five compounds are \textit{p}-type semiconductors. The Seebeck coefficient for metals is generally very low (a few~\textmu{}V/K), while for insulators it is very high. The values we obtained are characteristic of semiconductors that show definite promise for thermoelectric applications.
}

\subsubsection{Electrical conductivity}

\begin{figure}[h!tb]
\centering
\includegraphics[width=0.65\textwidth]{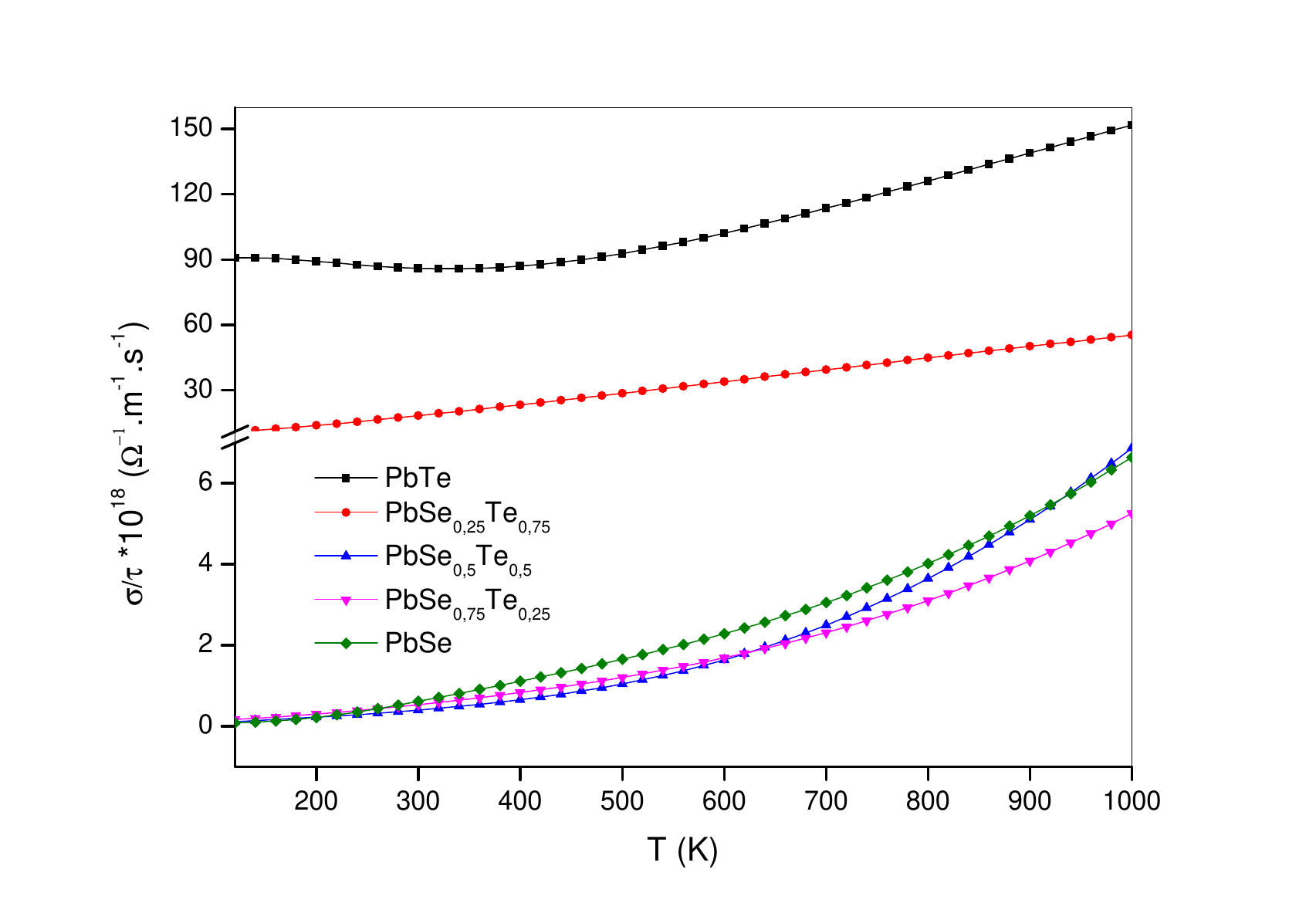}
\caption{(Colour online) Variation of electrical conductiviy versus temperature for PbSe$_{x}$Te$_{1-x}$alloys.}
\label{fig4}
\end{figure}

In figure \ref{fig4} we represent the variation of electrical conductivity over relaxation time as a function of temperature. {We can observe that the electrical conductivity of all five compounds increases with increasing temperature. The $\sigma/\tau$ values indicate significant differences in the quality of conduction in the materials, mainly due to variations in the concentration of $p$-type carriers. The binary compound PbTe exhibits the greatest values of electrical conductivity in the range of $10^{20}$ $\Omega^{-1}$ms$^{-1}$ which suggest the highest concentration of $p$-type charge carriers and therefore a better conduction. Next comes the $\textrm{PbSe}_{0.25} \textrm{Te}_{0.75}$ alloy with values of the order of $9\times10^{19}$ $\Omega^{-1}$ms$^{-1}$, where the conductivity is very close to that of PbTe, indicating that moderate incorporation of Se has a limited impact on the carrier concentration, then come the $\textrm{PbSe}_{0.5} \textrm{Te}_{0.5}$, $\textrm{PbSe}_{0.75} \textrm{Te}_{0.25}$ and PbSe with a lower conductivity (from $10^{17}$~$\Omega^{-1}$ms$^{-1}$ to $10^{18}$ $\Omega^{-1}$ms$^{-1}$). These selenide rich alloys have $\sigma/\tau$ values 100 to 1000 times lower than PbTe. This suggests a considerable decrease in $p$-type carrier concentration when Te is replaced by Se.
}

\subsubsection{Thermal conductivity}

When the electrical conductivity of a material increases, this necessarily implies an increase in its thermal conductivity, which is influenced by two phenomena: the movement of electrons and holes, known as electronic thermal conductivity ($\kappa_{\text{e}}$), and the vibrations of the crystalline lattice, known as lattice thermal conductivity ($\kappa_{\text{l}}$) \cite{Khan}, where the global thermal conductivity is given by equation \eqref{eqt5}:

\begin{equation} \label{eqt5}
    \kappa=\kappa_{\text{e}}+\kappa_{\text l}.
\end{equation}

To calculate the total thermal conductivity, we must first calculate the lattice conductivity. To do this, we used the Slack model formula \cite{Slack}
given by equation \eqref{eqt6}:

\begin{equation} \label{eqt6}
    \kappa_{\text{l}}=\frac{AM\Theta_D^3\delta}{\gamma^2Tn^{\frac{2}{3}}},
\end{equation}
where $A$ is a physical constant equal to:
\begin{equation}
    A=\frac{2.43\times10^{-18}\gamma^2}{\gamma^2-0.514\gamma+0.228}.
\end{equation}

$\Theta_D$, $\gamma$, $\delta^3$, $n$ and $M$ represents the Debye temperature, Gr\"uneisen parameter, the
volume per atom, the number of atoms in the primitive unit cell and the average mass of all the atoms in the crystal, respectively, where $\Theta_D$ and $\gamma$ were computed using the Gibbs2 software.

\begin{figure}[!t]
\centering
\includegraphics[width=0.7\textwidth]{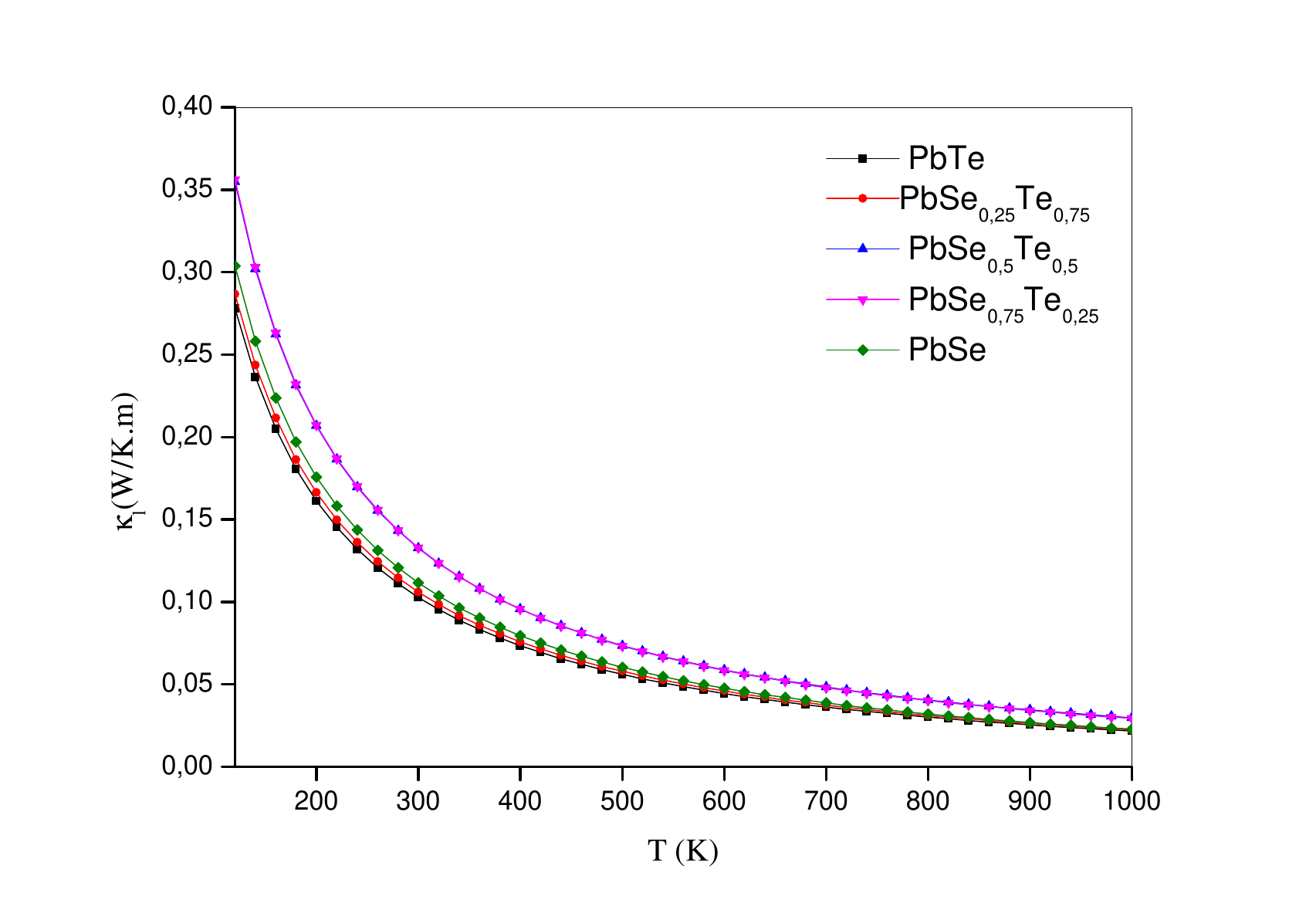}
\caption{Variation of lattice conductivity $\kappa_l$ versus temperature.}
\label{fig5}
\end{figure}

 Figure \ref{fig5} shows that the lattice conductivity decreases when temperature increases, and converge to a constant value at high temperatures.

\begin{figure}[h!tb]
\centering
\includegraphics[width=0.7\textwidth]{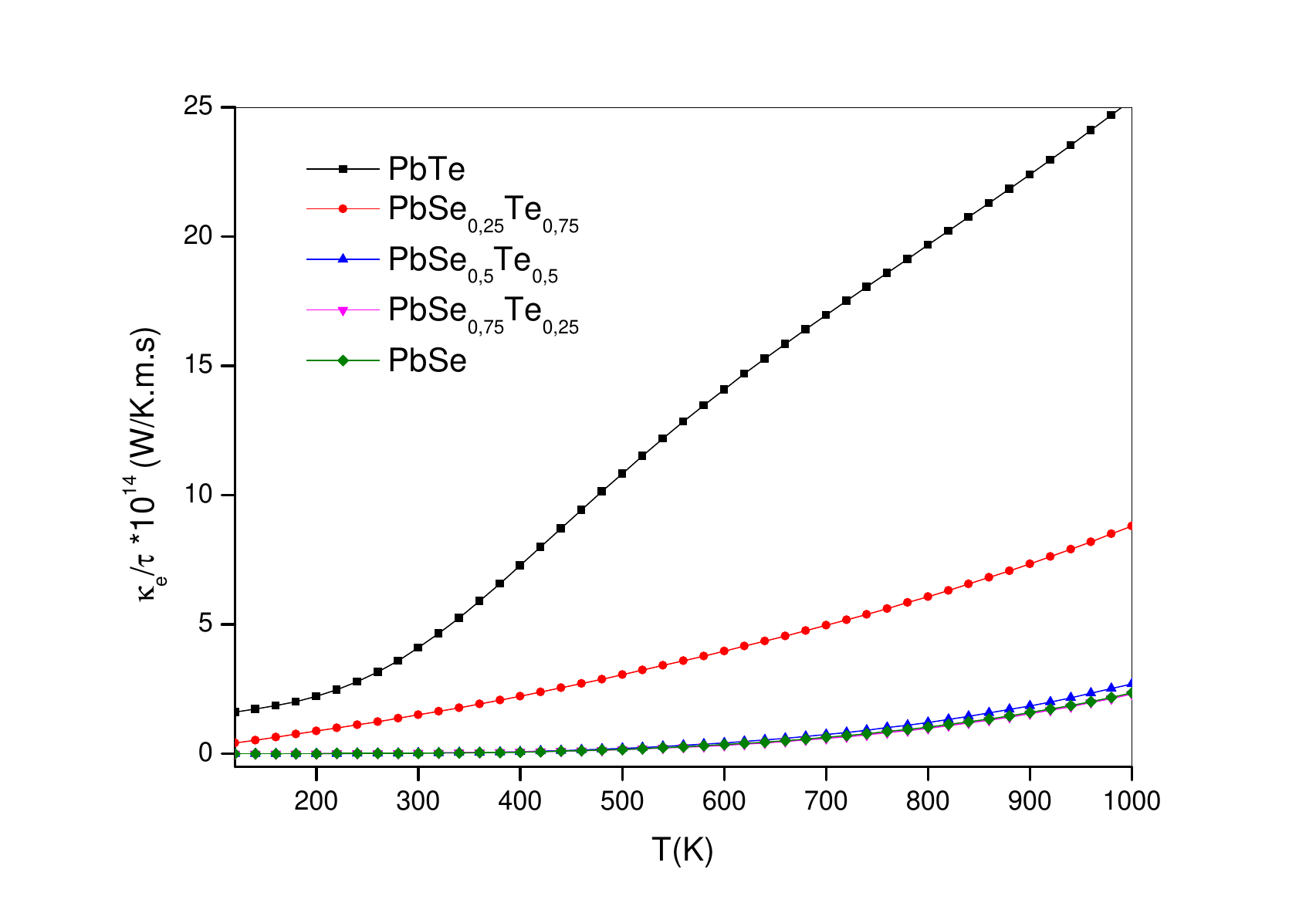}
\caption{Variation of electronic thermal conductiviy versus temperature.}
\label{fig6}
\end{figure}

 Figure \ref{fig6}, represents the variations of electronic thermal conductivity as a function of temperature. It is noteworthy that  $\textrm{PbSe}_{0.5} \textrm{Te}_{0.5}$, $\textrm{PbSe}_{0.75} \textrm{Te}_{0.25}$ and PbSe alloys have the lowest electronic thermal conductivity values (increasing from $2.82\times10^{11}$ $\rm{W}/$Km~s to $2.71\times10^{14}$ $\rm{W}/$Km~s). This coefficient for $\textrm{PbSe}_{0.25} \textrm{Te}_{0.75}$ alloys increases from $4.23\times10^{13}$ $\rm{W}/$Km~s to $8.81\times10^{14}$ $\rm{W}/$Km~s, and for PbTe it varies from $1.62\times10^{14}$ $\rm{W}/$Km~s to $25.3\times10^{15}$ $\rm{W}/$Km~s.

 {
The Selenium-rich alloys PbSe$_{x}$Te$_{1-x}$ with $x \geqslant 0.5$, exhibit the lowest electronic thermal conductivity $\kappa_{\text{e}}$, which is a highly favorable outcome for thermoelectric efficiency. This low $\kappa_{\text{e}}$ is consistent with their low electrical conductivity (low carrier concentration) but is counterbalanced by their excellent Seebeck coefficient ($S$ is high).
Ultimately, the Se-rich alloys, due to their high $S$ (as noted in section~\ref{sec3.1}) and very low $\kappa_{\text{e}}$ (as noted in section~\ref{sec3.3}), are the most promising candidates for achieving a high figure of merit \textit{ZT}, since they maximize the numerator $S$ and minimize the denominator $\kappa_{\text{e}}$ contributes to $\kappa_{\text{total}}$.
}

\subsubsection{Figure of merit}

The dimensionless parameter \textit{ZT}, which represents the figure of merit, is of crucial importance in thermoelectricity. It allows us to determine the efficiency of a material and whether it is a good candidate for thermoelectric applications.

\begin{figure}[!t]
\centering
\includegraphics[width=0.7\textwidth]{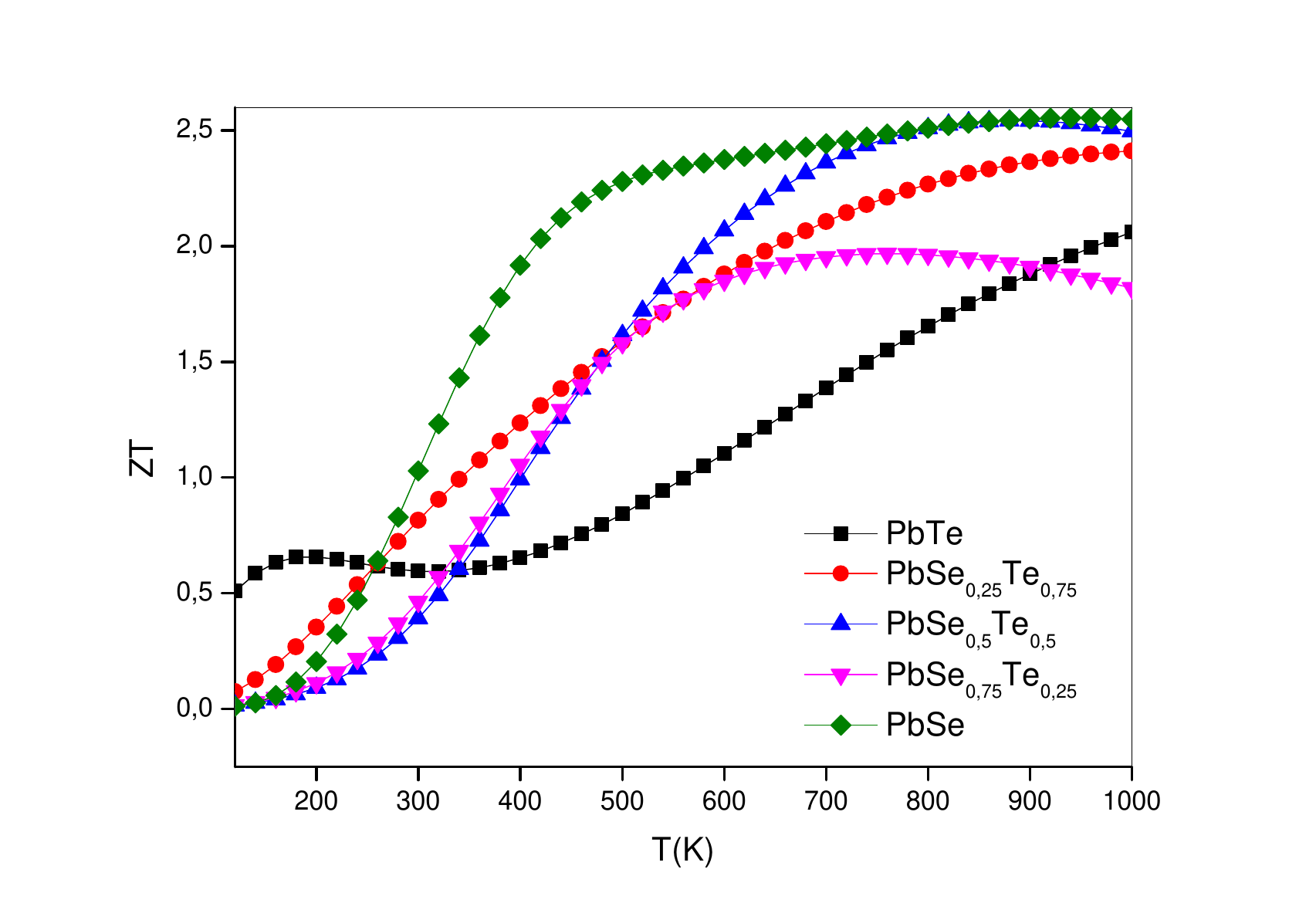}
\caption{(Colour online) Variation of figure of merit \textit{ZT} versus temperature.}
\label{fig7}
\end{figure}

To assess this, we plotted the variation of \textit{ZT} as a function of temperature in the range of 120~K to 1000~K for the PbSe$_{x}$Te$_{1-x}$ alloy compounds (\textit{x} = 0, 0.25, 0.5, 0.75 and 1) in  figure \ref{fig7}. As we can see, the value of this parameter increases with the increase of temperature. We can also see that we obtained a high value of \textit{ZT}, which
is due to a high Seebeck coefficient, high electrical conductivity and a low thermal conductivity. We can notice that the value of \textit{ZT} for PbTe, $\textrm{PbSe}_{0.25} \textrm{Te}_{0.75}$, $\textrm{PbSe}_{0.5} \textrm{Te}_{0.5}$, $\textrm{PbSe}_{0.75} \textrm{Te}_{0.25}$ and PbSe alloys reaches its maximum in the temperature
range between 800~K and 1000~K, with values of 2.06, 2.41, 2.54, 1.96, and 2.55, respectively. It is also worth noting that the PbSe alloy has the highest values of \textit{ZT}.

\section{Conclusion}

In this study, we conducted a numerical investigation using the DFT method to analyze the structural, electronic, and thermoelectric properties of the PbTe and PbSe compounds and their ternary alloys $\textrm{PbSe}_{x}\textrm{Te}_{1-x}$, with $x=$ 0, 0.25, 0.5, 0.75, and 1.

Structural properties were performed by using the generalised gradient approximation of Perdew Burke and Ernzenhof (GGA-PBE). Our results for the lattice parameter are in good agreement with the existing litterature on the subject, showing a non-linear variation with a slight bowing as a function of the composition \textit{x}.

Regarding the electronic properties, our compounds PbTe, $\textrm{PbSe}_{0.25} \textrm{Te}_{0.75}$,
%\newline
$\textrm{PbSe}_\textrm{0.5} \textrm{Te}_{0.5}$, $\textrm{PbSe}_{0.75} \textrm{Te}_{0.25}$ and PbSe exhibit a direct band gap $E(\textrm{L}-\textrm{L})$ with the values below 1~eV, indicating that they are direct band gap semiconductors.

Lastly, in the final section, we obtained promising results for the future of thermoelectricity. The materials we studied reached a value of \textit{ZT} equal to 2.06 for PbTe, 2.41 for $\textrm{PbSe}_{0.25} \textrm{Te}_{0.75}$, 2.54 for $\textrm{PbSe}_{0.5} \textrm{Te}_{0.5}$, 1.96 for $\textrm{PbSe}_{0.75} \textrm{Te}_{0.25}$, and 2.55 for PbSe, which is a highly interesting finding for future applications in thermoelectric devices.

\section*{Acknowledgement}
Thanks the financial support obtained from the University of Tlemcen through project, PRFU B00L02UN130120230001.

\bibliographystyle{cmpj}
\bibliography{cmpj_bib}

\ukrainianpart

\title
{Розрахунок термоелектричної конверсії у напівпровідникових сплавах PbSe$_{x}$Te$_{1-x}$}
\author
{М. Каїд Слімане, Б. Н. Брамі,
	М. Бушенакі, С. Бехечі } 
	\address{Лабораторія теоретичної фізики, факультет природничих наук, Університет Абу Бекр Белкаїд, Тлемсен, Алжир}

\makeukrtitle

\begin{abstract}
	\tolerance=3000%
Дане теоретичне дослідження зосереджено на структурних, електронних та термоелектричних властивостях PbTe, PbSe та їхніх потрійних сплавів PbSe$_{x}$Te$_{1-x}$ з використанням теорії функціоналу густини (DFT) методом повнопотенціальної лінеаризованої доповненої плоскої хвилі (FP-LAPW), реалізованого в рамках пакету Wien2k.
Структурні властивості визначені з використанням узагальненого градієнтного наближення схеми Пердью, Берка та Ернценгофа (GGA-PBE). Результати показують, що розраховані параметри ґратки добре узгоджуються з теоретичними даними, отриманими раніше. Щодо електронних властивостей, ми помітили, що для всіх сполук PbSe$_{x}$Te$_{1-x}$ існує пряма заборонена зона в точці $L$. Для термоелектричних властивостей ми використовували пакети BoltzTraP2 та Gibbs2. Наші результати показують, що сполуки PbSe$_{x}$Te$_{1-x}$ досягають значення коефіцієнта якості 2.55; це означає, що наш матеріал є хорошим термоелектричним зразком.
	\keywords свинець, метод DFT, напівпровідники IV-VI, електронні властивості, термоелектричні властивості
\end{abstract}

\end{document}